# Potentiality, Actuality and Non-Separability in Quantum and Classical Physics: *Res Potentiae* in the Macroscopic World

Robert R. Bishop[1]
Joseph E. Brenner[2]

*Abstract* To help resolve stubborn aporias in quantum mechanics, Kastner, Kauffman and Epperson have recently proposed a new interpretation of Heisenberg's ontological duality of *res extensa* and *res potentia*. In related work, Khrennikov and Aerts and their respective associates have explored the intermediate domain of mesoscopic phenomena in which quantum-like behavior can be observed and formalized. We generalize this work to define a *third* domain, one of macroscopic, interactive processes to which quantum concepts, but not quantum formalism apply. We show that they follow the non-standard logic proposed in the last century by the Franco-Romanian thinker Stéphane Lupasco. This logical system, grounded in the ontological dualities in nature, describes the evolution of complex processes and emergence at macroscopic levels of reality as chains of interacting actualities and potentialities. Without going outside the laws of physics, some new approaches to the problems of life, cognition and information can be made. The most appropriate formalism for describing this logic is under current scrutiny.

## 1. Introduction. The Hard Problems of Physics

The concepts of actuality and potentiality, and of the movement from the latter to the former, have been discussed since Aristotle, but now can be seen as common to both quantum and macroscopic levels of reality, that is, those of non-thermodynamic and thermodynamic change respectively. Areas of on-going debate in physics are on the one hand non-locality, wave function collapse, entanglement, *etc.* and on the other the possibility of extension of the relevant quantum mechanical formalism to selected complex macroscopic phenomena.

Quantum particles and fields are present at all levels of reality, but macroscopic objects do not display quantum properties such as non-locality, entanglement and spin, which are washed out by decoherence. Their behavior is determined by the residual electrostatic charge on assemblies of ions and inorganic and organic molecules. With the exception of some prepared physico-chemical systems, such as Bose condensates and some complex solids, the macroscopic classical and microscopic quantum domains do not, *prima facie*, share common properties. Some real process phenomena, especially human consciousness, are so complex that they appear to follow non-classical principles. However, attempts to explain consciousness by imputing localized quantum processes in the human brain, as well as the effect proposed by Conrad of sub-quantum fluctuations in biological processes have so far not been demonstrated satisfactorily.

In their recent article in this *Journal* [1], Kastner, Kauffman and Epperson (KKE) reinterpret Heisenberg's ontological duality of a *res extensa* and a *res potentia* as mutually implicative ontological extants that serve to explain some of the hard problems of physics. The authors state that quantum states instantiate a form of quantum *res potentiae* in a non-substance dualism as suggested by Heisenberg, the other term being a more or less standard *res extensa*. These quantum *potentiae*, however, are non-actuals and as such not spacetime objects "and they do not obey the (axioms of the) Law of the Excluded Middle (LEM) or the Principle of Non-Contradiction. … *Res extensae*, which constitute "structured elements of spacetime", are actuals that do obey LEM and PNC." This position, in our view, begs two important questions: 1) does the quantum physics of actuals in fact obey standard bivalent logic and 2) what logic, if any, might apply to quantum potentials during their 'long wait' for actualization by a measurement process?

[1] International Centre for Earth Simulation (ICES) Foundation, Geneva, Switzerland
[2] International Society for Studies of Information, Technical University of Vienna, Vienna, Austria (to whom correspondence should be addressed), joe.brenner@bluewin.ch

We agree that the Heisenberg picture can account for the counterintuitive features of quantum mechanics such as non-locality, *etc*. The realistic interpretation of KKE both provides a new explanation for these phenomena and supports the absence of a background spacetime, as proposed also by Rovelli [2]. However, a bivalent or multivalent truth-functional logic of propositions, of which the above axioms are a core part, could refer only to actual quantum entities as objects, assuming such could be identified. Quantum potentials are for us indeed ontologically real, 'pre-spacetime' possibilities but, as we will show, some potentials in macroscopic reality follow a logic of processes which is that of *quantum potentials* in that both the LEM and the PNC are abrogated.

The redefinition of the concept of potential in quantum mechanics has implications for the broader issue of the relation between the quantum and non-quantum domains. We do not call into question the principles of physical closure and completeness in quantum mechanics (QM). We state that a logic of processes is required to describe the evolution of complex interactive processes at biological, cognitive and social levels of reality. We propose that principles isomorphic to those of (QM) apply are necessary to describe the movement from actual to potential between pairs of dualities, ontological and epistemological, and *vice versa*.

Khrennikov [3], [4] and Aerts [5], [6] and their respective associates have provided a picture of an intermediate domain of macroscopic phenomena in which quantum-like behavior can be observed and formalized. We believe their approach is capable of further generalization to a domain of interactive processes to which quantum concepts, but not quantum formalism apply. We show that the latter are described by a non-standard logic of complex real processes proposed in the last century by the Franco-Romanian thinker Stéphane Lupasco. In the next Section, because of its unfamiliarity, a minimum presentation of this logical system will be made.

**2. Logic in Reality (LIR)**

In the period from 1935 to 1984, Lupasco proposed a logical-ontological theory, grounded in the quantum mechanics and physics of his time, of the origin and evolution of complex processes [7]. It has recently been up-dated and extended by one of us as Logic in Reality (LIR) [8]. In this paper, we show how some of the new interpretations of potentiality in QM can be positioned in relation to the principles of Lupasco's logic, in a sense confirming them.

All the real objects of the thermodynamic world are changing or have the potential for change. Complex thermodynamic relations and change, based on real potentials present in nature – chemical, electrical and cognitive – do not follow standard logics. (The existence of gravitational potential does not enter into our discussion directly.) Their evolution and their dynamic *qualitative* properties can best be described by a non-standard, non-propositional logic of processes, grounded in physics. The theoretical physicist Basarab Nicolescu, the continuator and friend of Lupasco, showed some thirty years ago that logical relations of isomorphism can exist between the laws governing phenomena at quantum and non-quantum levels [9].

LIR is a new *kind* of non-propositional logic grounded in the fundamental self-dualities and dualities in the nature of energy or its effective quantum field equivalent. These antagonistic dualities can be formalized as a structural, logical and metaphysical principle of opposition or contradiction instantiated in complex phenomena. The fundamental postulate of LIR, the Principle of Dynamic Opposition (PDO) is that all energetic phenomena (all phenomena) evolve, alternately and reciprocally, between degrees of actualization and of potentialization of themselves and their opposites or contradictions (A and non-A; Axiom of Conditional Contradiction) with which they are associated (Axiom of Functional Association) but without either going to the absolute limits of 0 or 100% (Axiom of Asymptoticity). The non-quantum phenomena we refer to here are all complex events, percepts, concepts, processes, theories, *etc.* in which there is a substantial degree of mutual determination by interacting dualistic elements. At the mid-point of maximum contradiction or interaction between A and non-A, a new third term or entity – T-state - can emerge at a higher level of reality or complexity - Axiom of the Included Third or Emergence). The Lupasco logic is thus a Logic of an *Included* Middle. This axiom is operative in the dynamic structure of the non-separable, contradictorial and inconsistent aspects of complex entities, processes and events at biological, cognitive and social levels of reality.

The LIR logical system has a formal part, axioms, semantics and calculus, and an interpreted part—a metaphysics and a categorial ontology. The semantics of LIR are non-truth-functional in the sense that their elements are not propositions at all, and the concept of truth-functionality (defined as valuations based on mappings between formulas and an algebra of truth functions defined on a given set of values, 0 or 1 in binary logic, several values in many-valued logics) should not be applied. The major aspect of the LIR 'dynamic' semantics is to give a sense of the dynamic state of the event, phenomenon, judgment, *etc.* whereby the event is moving between its actualization and the potentialization of its opposite. For the non-linguistic logical elements defined by the axioms of LIR, the dynamic, oppositional relation between two elements is expressed by implication: for any element **e**, **e** actual implies non-**e** potential. In LIR, the connectives of implication, conjunction and disjunction all correspond to real operators on the parameters of real elements. Accordingly, these operators are, also, subject to being actualized, potentialized or in a T-state[3]. They operate not on theoretical states-of-affairs or propositions, considered as the truth value of statements or measured values, but on property ascriptions of events, relations and processes. The calculus of LIR is based on the concatenation of implications of implications, an event calculus unlike any in the literature.

LIR is neither a physics nor a cosmology but a logic that enables stable patterns of inference to be made (in a mode close to abduction in the sense of Magnani), albeit with reference to metavariables that behave like non-Kolmogorovian probabilities in that the (abstract) limits of 0 and 1 are excluded. These limits are adequate only for simple physical situations in which there is no mutual transformation to all intents and purposes. As the analogue of the concept of observable, we might use that of *inferrable*. This term is already in use in a purely epistemic sense, but we use it as a designation of an energetic process element, 1) *what* is occurring, 2) *how* it is occurring, especially, the degree of interaction between the two process elements of interest and 3) the path by which 1) and 2) are inferred, that is conceived-perceived.

The values of logical values being probabilistic reflects the fact that for any real process in progress (being partly actualized and partly potentialized), there is a probability but no certainty that what is still potentialized will be actualized. Real entities can be described as encoding energy in potential form, as capacity for interaction. Both the actual and potential states of particles at lower levels—atoms, molecules, biopolymers, cells, *etc*.—are functional. The residual potentialities of entities at any level are the carriers of the information necessary for upward causation and emergence at the next higher level. This grounds, in basic physics, the concepts of "auto"-catalysis and "self"-organization in evolution and morphogenesis as not independent of context that includes opposites or antagonists, more or less actual or potential. For us, there is no 'self'-organization without 'hetero'-organization. The LIR logic of processes thus ascribes a role to potentiality in macroscopic systems without going beyond *standard* physics and mechanics.

---

[3] For a general discussion of operators in mathematics, logic and science, see Burgin and Brenner [10].

Logic in Reality is thus radically different from standard paraconsistent and paracomplete or intuitionist logics. In the former, the law of non-contradiction does not apply absolutely, but they remain linguistic logics concerned with the truth of propositions. In intuitionist logics, the law of the excluded middle is abrogated but only with respect to some formal aspects of mathematics. The law of non-contradiction is maintained. It is possible to approach a logic of quantum entities and their behavior from the point of view of propositions about them. The use of paraconsistent logic has been studied for this purpose by da Costa and Krause [11], in relation to the interpretation of the concept of complementarity between particle and wave properties introduced by Bohr. At present, we simply will outline the relation to quantum formalism that we think appropriate to the axiomatization of LIR. Applications of LIR have been made in the physics of information [12] and to perception and cognition [13]

**3. Quantum Logic**

Similarities exist between Logic in Reality and quantum logic. In 1968, Hilary Putnam wrote that quantum mechanics requires a revolution in our understanding of logic *per se*. "Logic is as empirical as geometry. … We live in a world with a non-classical logic." Quantum mechanics can be regarded as a non-classical probability calculus based on a non-classical propositional logic. Among other things, quantum logic focuses on the problem of the inability of measurement of two dependent quantities at the same time. Quantum mechanical states correspond to probability measures defined on an appropriate projection lattice of operators in a complex mathematical space (Hilbert space). This lattice requires a non-classical, non-Boolean logic for its description, which can be extended to other types of lattices, for example, of the properties of the system. In QM every probability-bearing proposition has the form "the value of physical quantity A lies in the range B". Aerts [14] has shown the mathematical structure that is constituted by these values can represent some of the properties of sufficiently complex physical macroscopic systems.

To show that the concept of a non-distributive projection lattice would be applicable to the LIR approach at macroscopic, in particular cognitive levels requires further work. However, it is no problem that a truth-functional semantics cannot be provided for the LIR connectives, given the notion of truth as reality that can be developed for LIR as a formal system. In any event, we have shown in our detailed comparison of LIR and other classical and non-classical logics that major changes in the meaning of their components, such as the connectives, must be made when moving to the extra-linguistic elements that are described by LIR.

According to Khrennikov [15], violation of the laws of classical probability theory is a statistical exhibition of violation of laws of classical Boolean logic. Thus, in logical terms the quantum-like modeling of cognition models non-classical reasoning, decision making, and problem solving. In particular, unconscious inference, including generation of a perception from an afferent sensation, is not based on the rules of classical logic. Quantum logic corresponding to the quantum formalism is just one special type of non-classical logic. In principle, there are no reasons to assume that human (mental) cognition, even if it has a non-Boolean structure, can be modeled completely with the aid of quantum logic and quantum probability. Still more general models might be explored. We consider that LIR is one such model.

In saying further that LIR is grounded in the *self*-dualities of the entities of the world, we refer to an (unexplored) relation to mathematics suggested by Majid and Heller [16], that no fundamental theory of physics is complete unless it is self-dual. We cannot offer a proof of this assertion, but simply take it as context for the evolution of processes, described by LIR, at higher levels of reality.

**4. The Quantum Formalisms of Aerts**

There is an on-going debate about whether a realist interpretation of quantum mechanics (QM), one which does not require a primitive notion of measurement, or an operational view that interprets QM as theory of measurement is to be preferred. We claim that the logic of/in reality is a quantum-type logic with the quantum probabilities of the quantum logical structure replaced by the statistically determined, inferrable, ascribable and in principle measurable values A and P of the alternating actualization and potentialization of dynamically contradictory states. As in probabilistic logic, the values also do not include the limits 0 and 1, but are reciprocally determined between greater than 1 and less than 0 (limits are only approached, asymptotically).

The formalism developed by Aerts [17] converts quantum mechanics into a system that can be applied to selected macroscopic phenomena, including space-time and the emergence of biological form and human cognition. The key point is that situations or entities that are intermediate between pure classical and pure quantum are not only possible, but their combined quantum and classical aspects can be described by different types of generalized mathematical structures. In this relatively quite new form of quantum logic, standard connectives themselves take on new, non-classical meaning, suggesting that, as in LIR, there is a close relationship between logic and quantum physics.

We wish to emphasize that despite the possible applicability of this formalism that we perceive, (for there exists no literature as yet), there are fundamental differences between Aerts' program and ours for which the reader should be prepared. The former makes no ontological commitment regarding the functional characteristics of the underlying particles and fields. Aerts' formal models are not directly concerned with being, that is the existence of the substrates that instantiate entities in real structures and processes. His theory is a theory of 'actions in the world' [18]. This approach does insist on the importance of the context in which an entity evolves, and the relationship between the entity and the context: it is referred to as context-driven actualization of potential (CAP) [19]. In the simple macroscopic systems chosen as examples, the Bell inequalities are violated, as they are by quantum particles that demonstrate non-locality. The interactions between system and context can thus be handled readily by the mathematical model. The criteria for analysis that develop from the application of the Bell inequalities allow an investigation of the probabilistic and logical structure of the available data. However, in talking about such violations of Bell's inequalities in language, Aerts says that not only are they not linked to contradiction, but "The contradiction *per se* is of no importance". Aerts thus excludes an essential role to what is defined in LIR as contradiction (or counter-action), that is, dialectic interaction, mutual change.

The concept of quantum morphogenesis, developed by Aerts [20] suggests a universal treatment of morphogenesis, understood as a temporarily stable change of form of both quantum and non-quantum systems, that does not depend on the details of the interactions that form a concrete ecosystem, organism or society. Systems are described by an abstract state-space, and the following aspects show the relation to LIR:

1. Sets of mutually inconsistent propositions are allowed, thus the law of non-contradiction does not hold absolutely. The situations involve non-Boolean logic and contexts, in which the logical value of the propositions depends on the history of the system. In LIR, the reciprocal relation between the degree of actuality and potentiality of a phenomenon and its contradiction in the principle of antagonism are such 'propositions'.

2. The systems are probabilistic. Morphogenesis is described in terms of probabilities or uncertainties associated with given sets of propositions. The contextual nature of the propositions requires non-classical probability distributions (non-Kolmogorovian). In LIR, logical values are also contextual, i.e., also depend on the history of the system (are systems of systems, etc.), and the shifts from actual to potential and inversely are probabilistic.

3. Feedback is a crucial element. Changes in the environment and system interact and influence one another. In LIR, all complex systems involve feedback, enabling a parallel with Aerts' construction.
    Aerts' key point is the following:
"What makes our construction essentially different from the models one finds in the literature is the role of non-commutativity of the system of propositions. Non-commutative propositions are related by uncertainty principles and are typical of systems which cannot, without an essential destruction, be separated into independent parts." Aerts hoped that his "quantum mechanical model for the cognitive layer of reality could be an inspiration for the development of a general interactive logic that could take into account more subtle dynamical and contextual influences than just those of the cognitive person on the truth behavior of cognitive entities." This is what we propose LIR is in principle capable of doing, but using the dynamic actual and potential states of processes instead of propositions about them.
    In the key 2012 summary of Aerts' work, "Potentiality States: Quantum versus Classical Emergence [5], Aerts and D'Hooghe listed the three most studied examples of the appearance of potentiality states in macroscopic *cognitive* reality, indicating the presence of 'Hilbert space-like' features. Unfortunately, all of these examples refer to *binary* aspects of reality: vessels full/empty of water; the Liar Paradox: yes/no states; the undetermined opponent, A/B of a soccer team. Aerts' analysis is correct, but it is difficult to extend it to non-cognitive or complex cognitive phenomena.
    We seek the operation of our Logic in Reality elsewhere, in the emergence of 'team spirit', in dualistic interactions in biology, in the instantiation of absence as discussed in relation to information, in the relation between consciousness and unconsciousness. Further contrasts between our program and that of Aerts will appear in what follows.

**5. Non-Separability. Systems**
    Many physical as well as philosophical arguments depend on some form of absolute separability of dichotomous terms *via* the importation, explicit or implicit, of abstract principles of binary logic exemplified in the use of standard notions of time, space and causality. Standard category theory requires that principles of exclusivity and exhaustivity. In contrast, the critical categorial feature of the LIR process ontology is the *non-separability* of opposing phenomena (two theories) or elements of phenomena, *e.g.*, syntax and semantics, types and tokens.

As discussed by Healey [21], a physics treats systems by assigning them energetic states, and non-separability as a physical principle, functioning at not only quantum but also at biological and cognitive levels is related to holism. Holism is the thesis that the whole is more than the sum of the parts, and non-separability can be defined by the statement that the state of the whole is not fully constituted by the states, properties and relations of the parts. These do not provide the complete basis for the whole, and one says in this case that the whole — an object or process with its own set of properties and relations — does not supervene on its parts. Since we are talking about states of systems, we use Healey's principle of state separability: the state assigned to a compound physical system at any time is supervenient on the states then assigned to its component subsystems (the latter are the basis for the former). In this connection we should emphasize that in LIR, systems are constituted and persist due to the balanced operation of real forces, attractive and repulsive, necessary for the their formation and persistence. Systems can be both epistemological and ontological entities, but in both cases changing interactions are involved that cannot be described by a predicate logic of sentences.

A further ascription of non-separability to classical processes such as phase change and the propagation of gravitational energy is possible, but our concern is with processes that are more complex due to their embodiment of mutual interaction. Healey said that: "Non-separability would be a trivial notion if no qualitative intrinsic physical properties were ever assigned at space-time points or in their neighborhoods. But this would require a thorough-going relationism that took not only geometric but all local features to be irreducibly *relational*." This is a 'relationalist' thesis to which LIR subscribes, like that of Relational Quantum Mechanics [22]. The local features are the energetic states of actualization and potentialization and the emergence of new entities. Quantum entanglement is the paradigm example of non-separability, but so can be the emotional-social relation between pairs of human beings.

The '*Separability Principle*' formulated by Karakostas [23] is that the states of any spatio-temporally separated subsystems S1, S2, ..., SN of a compound system S are individually well-defined and the states of the compound system are wholly and completely determined by them and their physical interactions including their spatio-temporal relations. The phenomenon of quantum non-separability reveals the holistic character of entangled quantum systems. Quantum mechanics is the first and till today the only mathematically formulated and empirically well-confirmed theory, which incorporates as its basic feature that the 'whole' is, in a non-trivial way, more than the sum of its 'parts' including their spatio-temporal relations and physical interactions. Contrary to the situation in classical physics, when considering an entangled compound system, 'whole' and 'parts' are related in such a way that their *bi-directional reduction* is, in principle, impossible. This formulation is suggestive, and it ties into the mereology defined by the Lupasco of a dynamic relation (dialectics) between parts and wholes in which each share some of the other's properties, following the PDO.

## 6. Energy, Potentiality and Actuality

As discussed in [8], the juxtaposition of the terms energy, actuality and potentiality goes back to Aristotle. The condition of an entity whose essence is fully realized is an entelechy, a condition of actuality as distinguished from potentiality. Energy is derived from a verb, *energein*, and implies duration in the operation of the driving force of the process, whereas entelechy refers to its two crucial instants, namely, the start of the movement or conception of a project and its achievement (*telos*).

The significant contribution of Lupasco was to see the link between 1) the actuality and potentiality of Aristotle and the other classical dualities; 2) the actual and potential energy defined by the early 20th Century energetists; and 3) the new (in 1925) Heisenberg uncertainty principle. The principle of dynamic opposition, Lupasco thought, required redefinition of the logical underpinnings of all aspects of philosophy, metaphysics and ontology. As discussed by KKE, Heisenberg considered that the probability wave of quantum mechanics was a *potentia*, a new kind of physical reality "halfway between the massive reality of matter and the intellectual reality of the idea", and the reduction of the probability wave during measurement was a movement from potential to actual. His intuition that the choice of the term *potentia* implies is intriguing, as is the 'halfway' that reminds one of the Lupasco T-state.

Kastner *et al.* start by defining the *res potentia* as a 'substance' in a general, Aristotelian sense, the essence and definition of a thing, "one of the two fundamental, mutually implicative ontological constituents of nature at the quantum mechanical level." But it can also be understood (KKE go on to say) 1) as a general concept applying to a broad range of possibilities; and 2) a subset of those possibilities constituted by quantum entities and processes, which provide "strong candidates for realism" – quantum *potentiae* (QP). The further description of the relation between these two terms is that a quantum entity, before it is actualized as a *res extensa* is a non-local quantum *potentia.* From this 'non-substance duality', measurement can generate acausally a new actual which can in turn bring about new quantum possibles.

The work of KKE is thus essential to counter the on-going lack, in the literature, of ontological commitment as to the role of potentiality, going back to Russell's statement that logical constructions (in the standard sense) are to be substituted for inferred entities. No reference that we have uncovered suggests that non-localized potentialities not only have real existence, fitting the *logical* foundations of LIR, but also are *functionally* related to actualities, an Axiom in LIR. If one follows this line, one arrives at potentiality as an inferred physical entity.

Aerts [24] stated, supporting the LIR picture, that "*Change is described by potential properties becoming actual and actual properties becoming potential* (emphasis ours)." The reciprocity of the interaction between actual and potential remains is central to LIR. What drives the change or 'becoming' is the overall energy gradient of the universe, but what relates the two is the dynamic opposition inherent in energy and all of its manifestations.

**5.1 The Transactional Interpretation of Quantum Mechanics**

The Transactional Interpretation of quantum mechanics (QM) which Kastner and Cramer present in [25] and [26] describes an interaction that is similar in *form* to the essential generalized dualistic interactions which we see taking place routinely at the macroscopic level. The Possibilist (not probable, pre-probability) Transactional Interpretation of quantum mechanics proposes a 'growing universe' picture, in which actualized transactions are the processes by which spacetime events are created from a substratum of quantum possibilities. Again, we argue not for the existence of such a substratum at macroscopic levels, but one of non-quantum actualities and potentialities as the basis and substance of real change.

For QM, the basic transactional picture involves an 'emitter', an 'absorber' and the clearly defined response of the absorber, the latter being normally excluded from the analysis. The absorber response is the key component of the non-unitary measurement transition, but processes are still indeterministic. Processes go from actualized potentialities, described by a Kolmogorovian probability space, here termed incipient, to actualities. In LIR at the macroscopic level, the same is true but the probability space (we claim) is non-Kolmogorovian, and determinism enters with thermodynamics. For us, the boundary between thermodynamic and non-thermodynamic is 'where' a hydrogen atom loses an electron to become an ion. In QM, the quantum vector potential consists of two terms, both required for the reality of the field, which are complex conjugates of one another, reflecting the "essence of the transactional process". This is an equally valid description of the two terms of the principle of dynamic opposition in LIR. To repeat, since the processes discussed by LIR are macroscopic, in or better accompanied by spacetime (see next Section 6), new entities can emerge from a point of maximum energy of interaction. On this basis, the formal structure of non-quantum interactions would seem a profitable area for further work.

In Logic in Reality, the spacetime realm is that not only of concrete, actualized events but of events jointly constituted by actualities and potentialities in a dynamic interplay. This realist statement supports the realism of the extension by Kastner of the Transactional Interpretation and its extensions to the relativistic domain [26], which look like complex macroscopic reality as conceived of by Khrennikov. Real spacetime events always involve the transfer, rather than emission and absorption of real energy (or perhaps better 4-momentum) and define a spacetime and a principal temporal direction.

## 6. Spacetime

We feel that the calls by Kastner, Zeilinger and others for a reconceptualization of the usual notions of time and space have been in part answered in the logical system of Lupasco. The joint quantum-non-quantum perspective that we are trying to develop in this paper is, among other things, tolerant of expressions dealing with spacetime such as 'pre'-spacetime possibilities, 'before' spacetime or event. Of the theory proposed by Lupasco [7], we can give here just three essential theorems and a brief discussion:

**1: The actualization or potentialization of a logical event is not a function of time, but time that is a function of the dynamics of actualization and potentialization.**

If an actualization of an element or its opposite is rigorous and absolute, there is no more time; the logical element is fixed and immutable. Process as such is impossible. If the consequent potentialization is, accordingly, infinite, the element disappears along with the temporality. The notion of time enters into the concepts of wear, change and transformation, all of which require modification of identity. Complete actualization or potentialization would be equivalent to an end of time, in the heat death of the universe, for example, the absolute homogenization of energy at the lowest level. Time is only possible due to the existence of contradictory dualities whose energetic antagonism is both the source and necessary condition of partial, non-infinite actualizations and potentializations.

**2: Objects and events do not exist or take place in time, but are the sources of, or 'unroll', (*déroulent*) their own time.**

According to LIR, there are three kinds of time: a positive time corresponding to the identifying actualizations of physico-chemical causality; a negative time inverse to the former of differentiating actualizations, associated with the processes of living matter. The two involve both continuities and discontinuities, and their interaction results in the emergence of a third time, at the mid-point, corresponding to a minimum of non-contradiction and a maximum of tendency to contradiction. This is the time of quantum and neuropsychical entities, a logical basis for the phenomenological 'nowness' of Varela.

**3: Objects and events do not exist or take place in space, but are the sources of, or 'unroll', (*déroulent*) their own space.**

Thus objects are not in space, but space is in objects; objects are not localized, but localize, create localizations. It is in this admittedly informal way that such a space has the same characteristics as a configuration space, that is, it is a function of the number of its elements and of their degrees of freedom; it is what links the elements, their relations, that permits their co-existence in a system and their simultaneity. We thus construct three (kinds of) space: 1) a positive space of the physical world and its matter, of particles following Bose-Einstein statistics and by analogy that of the set M of the Axiom of Choice of Zermelo-Frankel set theory, a space that could be called photonic space; 2) a space of biological configurations, particles following Fermi-Dirac statistics, the sets N of choice, or electronic space; 3) the third is the space of interactive quantum phenomena and of esthetic and psychological phenomena as well as of the sub-sets P of the Axiom of Choice. There is no spatial location outside of what is inside it. Logical space and logical time constitute a space-time proper to each system, a configuration space-time.

It is therefore gratifying to read in Kastner [27] that spacetime is not a substantive manifold that becomes occupied with events and objects; rather, spacetime itself exists only in virtue of specific actualized events. This implies for Kastner that spacetime is discrete rather than continuous, and that properties attributed to spacetime based on the notion of a continuum are idealizations that do not apply to the real physical world. LIR, however, accepts the idea that the macroscopic world does display phenomena that are continuous to all intents and purposes and there is no reason, again, not to accept that both exist. Kastner suggests that "Spacetime itself cannot be above or beyond such considerations, namely, quantum randomness and quantum entanglement" but, we add, not beyond the properties of the existence of 'anything at all' and what it implies. Clearly, what is missing in the LIR approach are some more operative links to mathematical physics, in order to see what kind of 'space' (Fock space?) the potentialities and actualities of real processes evolve.

**7. Other Relevant Aspects of Quantum Mechanics**
**7.1 Ontic Models. Khrennikov Supplementarity**

The above structure proposed by Lupasco for spacetime can be related to the concrete ontic model preceding quantum mechanics (the latter is treated as an epistemic model), namely, the prequantum classical statistical field theory (PCSFT) proposed by Khrennikov [28]. PCSFT yields a natural physical interpretation for the basic quantum mechanical entity - the quantum state ("wave function"). We propose that the Lupasco approach is an ontic model in this sense.

The Lupasco theory does not exclude pure quantum states *a priori* nor simple macroscopic states. It calls for the existence of what is usually called 'mixed' states but which we prefer to designate as interactive states instantiating actuality and potentiality, for example in what Khrennikov calls the ontic-epistemic correspondence.

A main objection to ontic models in the standard view is that since ontic entities are unapproachable by our measurement devices, it is totally meaningless to put efforts to theorizing which does not lead to explicit experimental consequences. We agree with Khrennikov's short answer is that by gaining the knowledge about nature we use not only outputs of observations (including our senses), but also the power of our mind, both logical and transcendental reasoning.

This can lead us to laws of nature which are hidden in observational data. The Lupasco Principle of Dynamic Opposition is one such law.

It is also noteworthy that Khrennikov has no requirement for a classical-quantum dichotomy: "One of the main characteristics of the ontic state is the possibility to assign to it the concrete energy. Here energy is treated totally classically, as in classical field theory. Probabilistic ontic states are given by probability distributions or by random fields, *i.e.*, random variables. Such random variables can be expressed as a function of a chance variable and a space variable, and this scheme is easily extended to the case of the abstract Hilbert space H. Here ontic states are vectors (in general non-normalized) belonging to H. Instead of the energy density, "energy at a fixed point of space", one can consider the "energy along some direction", a vectorial concept. Probabilistic ontic states are H-valued random variables, i.e., random H-fields. For such a state, we can find the average of its energy along a direction. The space of probabilistic ontic states R can be mapped to the space of epistemic states D, each epistemic state is the image of some class of probabilistic ontic states. PCSFT provides the new viewpoint on superposition of quantum states - by treating them as images of prequantum random fields."

Appealing to the special ontic model of the classical random field type (PCSFT), clarifies the meaning of a quantum state (in particular, a pure quantum state - "wave function"), resolving the contradiction between interpretations based on "physical waves" and "waves of probability". There are "physical waves", mathematically represented as classical random fields, *e.g.*, photons are just classical electromagnetic fields. Such sub-quantum fields propagate causally in physical space. However, the wave function is not the ontic wave. The wave function and more generally the quantum state are covariance operators of sub-quantum random fields. Thus the wave function can be interpreted probabilistically as the "wave of correlations." We claim that the movements from potentiality to actuality and the reverse can be seen as ontic 'waves' or concatenations of states, but the formalism is lacking to support this claim.

Over ten years ago [29], Khrennikov proposed a principle of *supplementarity* to go beyond the somewhat idealized relations of Bohr complementarity. Without going into detail here, it says that quantum probabilistic behavior need not imply the impossibility of constructing a realistic underlying model. A consequence of the principle of supplementarity which could stimulate experimental research is that quantum-like probabilistic effects might be observed for systems and contexts (e.g. physical or biological) for which existence of an underlying realistic model looks very natural. The observation of interference of probabilities would not be considered as a contradiction with the realistic model. Such effects might be observed for ensembles of macroscopic physical systems or for ensembles of cognitive systems. These are exactly the systems whose behavior is described by Logic in Reality.

**7.2 Quantum Non-Locality**

A recent paper by Esfeld [30] illustrates the Lupasco approach, and its relation to KKE and the work of Khrennikov in several ways. Esfeld defines a version of ontic structural realism (OSR), following Bohm, in which particles, matter or 'flashes' of energy at points in physical space, in the primitive ontology of QM, constitute a structure represented by the wave-function. "That structure is a concrete physical entity in spacetime consisting of a network of relations of entanglement encompassing all these elements." QM, in contrast to classical mechanics, involves 'some sort' of holism, non-separability (see above) or structuralism expressed by the entanglement of the wave-function. Non-locality means that due to entanglement of the wave-function, quantum systems are not (totally) separate individuals, but their temporal development occurs in tandem, whatever their spatial distance may be.

Esfeld claims that his decisive point is that the primitive ontology theories of QM and OSR can 'help each other' - the former providing an ontological status and structure of the wave-function while QM provides the concrete physical entities in physical space that instantiate the structure. This is an example, for us, of a macroscopic interaction, at a cognitive level, of theories following the rules of LIR, as well as being a restatement of the underlying non-separability of dualistic phenomena.

**7.3 Potentialities to Actualities**

In related work [4], Conte, Khrennikov and Zbilut have recognized the importance of the transition from states of potentiality to their actualization as the "basic mechanism of our reality". Aristotle is given credit to this and, also, *that actualities are the origin of potentialities* which can generate new actualities. The authors make the additional, for us critical statement, citing Aerts, that in quantum systems forms of potentiality coexist with forms of actuality in the Aerts' description of evolution. "Matter must be considered in its potential form in addition to its actual form."

Conte *et al.* assign a central role to spin as an expression of "a more general and differentiated essence and modality of self-fulfillment of reality at its various levels", beyond the usual one in QM and find a mathematical proof that the transition from potentiality to actuality is central to it. What we have reservations about is the subsequent reference to transition from 'pure' potentiality to an actuality, which is in our view true *only* in QM. Supporting our critique of this point, the authors claim that that their "algebra recovers the first two principles of logic that are the Principle of Non-Contradiction and the Principle of the Excluded Middle". In other words, they come to exactly the same conclusion as KKE (cf. Section 1).

*We* conclude that what are extracted here are classical properties of quantum objects. This discovery does not and should not exclude the existence of non-classical or quantum-like properties of macroscopic objects or processes, and indeed Khrennikov has identified some of them. In [28], he specifically replaces the language of propositions which is common in quantum logic with a language of wave functions, proposing a quantum-like representation algorithm (QLRA) which can map probabilistic data and represent "hidden macroscopic configurations" in the natural sciences. He finds that all distinguishing features of quantum probabilistic behavior can be modeled by using macroscopic systems, deduced from a system of natural probabilistic axioms. However, for such macroscopic models the QLRA is not complete; hidden variables are required, not observable because too "fuzzy".

This is where our logical approach enters the discussion: the variables, which we refer to as the reciprocally determined values of potentiality and actuality, are not observed or calculated but they and their evolution or change can be inferred. The status is one of less certainty than that of mathematical proof but has the minimum rigor of a naturalized metaphysics, also starting from "natural probabilistic axioms". From this standpoint, we can agree with Khrennikov's statement to the effect that contexts are the fundamental elements of any probabilistic model with the indicated properties.

In LIR, since no individual term is an identity, that is, unconnected to other terms, one has the same relation as that between a term and the context that perturbs it. Neither the commutative law of standard logic nor the distributive law between conjunction and disjunction holds. Any applicable formalism is, accordingly, non-Abelian and non-Boolean respectively, and the resulting probability distributions are non-Kolmogorovian.

We believe that the real potentialities and actualities, the logical elements with the T-state defined by the Lupasco axioms, constitute the quantum-like "supplementary observables" proposed by Khrennikov which violate classical probabilistic structure. They should be capable of being represented in a quantum-like way by non-commutative self-adjoint operators or some other formalism which remains to be defined, but whose principles we consider established.

In 2012 [31], Darvas showed that standard causality, requiring invariance under time-reflection, works in flat space-time, but this is not the case in real non-classical physical situations. Causal paradoxes in QM are no longer "paradoxes" but normal phenomena in nature; that is, in the logical framework where they can be interpreted in the domain of the General Theory of Relativity, causality really does not work, at least in its traditional sense. One has to give up causality in both QM and classicality. The described reconsideration of causality occurs, for example, in respect to the interpretation of local theories. The original approach of Lupasco to causality, as one such 'local theory', should be worth further debate (for details see [8]).

## 8. Possibilities, Potentiality and Probabilities

Like KKE, Karakostas [23] also related the phenomenon of quantum non-separability to potentiality; it effectively provides a coherent picture of the puzzling entangled correlations among spatially separated systems. A generalized phenomenon of quantum non-separability further implies contextuality for the production of well-defined events in the quantum domain, and contextuality entails in turn a structural-relational conception of quantum objects, viewed as carriers of dispositional properties. Contextuality naturally leads to a separable concept of reality whose elements are experienced as distinct, well-localized objects having determinate properties.

In the general Lupasco scheme, the terms of possibility and potentiality define domains to which binary and ternary logics apply respectively, that is, the former does not involve dynamic interactions, and the latter does [32]. Logic in Reality offers a definition of probability *vs.* possibility that clarifies some previous work by Karakostas which is of interest here because he also cites the account of Heisenberg as characteristic of the significance of the notion of potentiality as a clarifying interpretative concept for quantum theory. He writes: "The probability function ... contains statements about possibilities or better tendencies ('*potentia*' in Aristotelian philosophy), and these statements ... do not depend on any observer. ... The physicists then speak of a 'pure case'. [Consequently,] it is no longer the objective events but rather the probabilities for the occurrence of certain events that can be stated in mathematical formulae. It is no longer the actual happening itself but rather the possibility of its happening, the *potentia*, that is subject to strict natural laws. ... If we ... assume that in the future exact science will include in its foundation of the concept of probability or possibility the notion of potentia, then a number of problems from the philosophy of earlier ages appear in a new light, and conversely, the understanding of quantum theory can be deepened by a study of these earlier approaches to the question". For Lupasco, that something *s* is possible implies only its own negation, that of the impossibility of it happening (not the negation of possibility "it is not possible that *s*"). An element in a predominantly potential state does not *imply* its non-actualization. The actualization may not occur, but it would require an input of energy, *via* an accident or event, that is extrinsic and unpredictable, even if deterministic.

The possible involves a random choice without any determinism or energetic capacity, a disjunction between a yes and a no, without an antagonistic 'partner'. (Epistemic possibility, what one knows about a possibility, is context-dependent and shades over into probability. This concept does not affect the distinction made here, since the set of binary choices is still the only one available.). In the possible, contradiction disappears in the yes or no as isolated states, that is, in pure non-contradiction. The potential, on the other hand, contains or is always accompanied by the actual - that which opposes it and prevents it from becoming actual or actualizing itself. Potentiality thus implies a rigorous determinism, which is not found in the possible.

At any point in time, a dynamic phenomenon will be actualized and potentialized to a certain, probabilistically determined degree. The key point is that the sum of these degrees must be greater than zero but less than 1, since complete final states cannot be achieved by complex process entities.

## 9. *Res Potentiae* in the Macroscopic World. Living Systems

For purposes of discussion, we may say that the cases studied by Aerts *et al*. [5] (cf. Section 5) are 'prepared systems', like those of single particles in a physical trap, or single cognitive notions in a 'temporal trap'. They are well described by the quantum formalism indicated but their implications for the everyday world are not clear. In contrast, the Lupasco logical system allows inferences to be made about all real complex systems, processes and behavior, recognizing that they will range in purport from trivial to significant.

Giuseppe Longo [33] pictures a unifying theoretical framework for biology that is not based the specific invariants and invariants preserving transformations (symmetries) of (mathematical/theoretical) physics. Such a framework should focus on the permanent qualitative, non-generic changes that modify the analysis of processes both in ontogenesis and evolution. These biological changes preserve an ever changing structural stability, the coherence of organisms. The adaptivity of an organism and the diversity of a population are consequences of variability, thus of randomness. They contribute in an essential way to the stability of life phenomena. Thus, in a sense, variability may be considered as one primary invariant of the living state of matter.

Longo analyzes organisms with the tentative notion of extended criticality, where the notion of criticality is extended from being pointwise in physics to being relevant for a whole region of the description space. Actual biological phase space (functions, phenotypes, organisms) in our view can be similarly described by the axiomatics of LIR. Lupasco demonstrated the applicability of his principle (PDO) governing the operation of potentialities and actualities biological entities from proteins to organisms [34]. In Longo's terms, biological possibilities, better probabilities, are the result of an unpredictable sequence of symmetry changes. This situation is in contrast to the invariant (conservation) properties which determine physical "trajectories", in the broad sense (including for Hilbert spaces).

By the lack of mathematically stable invariants (stable symmetries), there are no laws which entail, as in physics, the biological observables in the becoming of the biosphere. In physics, the geodetic principle mathematically forces objects "never to go wrong". Living entities, instead, may and in fact do follow many possible paths (Lupasco: paradialectics), and they go wrong most of the time. Longo's thesis is that evolution and ontogenesis are "diachronic processes" of becoming that "enable" the future state of affairs and do not cause it in the physical sense... Instead, such entailed causal relations must be enriched by "enablement" relations for biological processes. Physical quantities typically play a different role in biology than in physics. In biology, echoing Stuart Kauffman, Longo considers that they play the role of constraints, limiting possibilities on the one side and enabling behaviors on the other side. A description should be possible of the qualitative evolution of life.

## 10. Conclusion and Outlook

The deepest new insights into the structures of our world are coming from quantum physics and astrophysics. However, there is no reason not to take into account insights gained from observation of processes at the human level of reality, especially where these display a congruence or isomorphism with quantum phenomena, thus reproducing some of their foundational principles. To insure a satisfactory degree of rigor in the absence of quantification, some form of logic would seem appropriate that applies in the classical domain but whose ontology does not *a priori* eliminate the phenomena of interest in the absence (till now) of an appropriate mathematical form.

This paper has proposed a candidate for such a logic. Its methodology is primarily one of implication and logical inference and its domain of application, to repeat, that of non-quantitative change in biological, cognitive and social processes. In our logical context, the existence of *res potentiae* in macroscopic processes permits inferences about their properties and their probable evolution. This concept may be considered further evidence for the universality of the principles governing the quantum domain, percolating as it were into the classical domain which is grounded on it. It can be related to current issues in computation and information science and philosophy. To explain the evolution of complex macroscopic processes with such quasi-quantum-like behavior, Logic in Reality does the work of mathematics. Its finality would be the naturalization, the bringing into science, of relational domains of human existence – values, opinions, attitudes, behavior – which have been excluded due to their non-conformity with the logical and categorial criteria for science as exact.

We thus have taken seriously the call of Kastner, Kauffman and Epperson (KKE) for a re-examination of standard notions of spacetime as the extant domain in relation to non-quantum as well as quantum phenomena. We note that theories of non-quantum phenomena which make use of quantum formalism do so in a highly selective manner. Our view starts with a reinterpretation of the logical-dynamic relations in general macroscopic ontological change, noting their isomorphism with *some* foundational features of the quantum world, in particular Heisenberg's conception of potentialities (*res potentiae*) as presented by KKE . Our demonstration that the relation of potentiality to actuality in the quantum and non-quantum domains is one of isomorphism could be a contribution to the basic physics of both.